\documentclass[amsmath,amssymb,twocolumn,superscriptaddress]{revtex4-2}

\usepackage{graphicx, bm, physics, MnSymbol}

\usepackage{soul}
\usepackage{xspace}
\usepackage[dvipsnames]{xcolor}
\usepackage[normalem]{ulem}
\usepackage{mathrsfs}

\usepackage{url}
\usepackage{hyperref}
\usepackage[htt]{hyphenat}
\hypersetup{
  colorlinks   = true, 
  urlcolor     = blue, 
  linkcolor    = blue, 
  citecolor   = red 
}

\newcommand{\kket}[1]{|{#1}\rrangle}
\newcommand{\bbra}[1]{\llangle{#1}|}
\newcommand\mc{\mathcal}

\newcommand{\hmc}[1]{\hat{\mathcal{#1}}}
\newcommand{\olg}{\overline{\gamma}}

\newcommand{\ops}{\mathscr{L}}
\newcommand{\omc}[1]{\ops(\mathcal{#1})}
\newcommand{\hms}[1]{\hat{\mathscr{#1}}}

\begin{document}
\title{Uniformly Decaying Subspaces for Error Mitigated Quantum Computation} 
\author{Nishchay Suri}
\email{nsuri@usra.edu}
\affiliation{QuAIL, NASA Ames Research Center, Moffett Field, California 94035, USA}
\affiliation{USRA Research Institute for Advanced Computer Science (RIACS), Mountain View, California 94043, USA}
\author{Jason Saied}
\affiliation{QuAIL, NASA Ames Research Center, Moffett Field, California 94035, USA}
\author{Davide Venturelli}
\affiliation{QuAIL, NASA Ames Research Center, Moffett Field, California 94035, USA}
\affiliation{USRA Research Institute for Advanced Computer Science (RIACS), Mountain View, California 94043, USA}
\date{\today}

\begin{abstract}
We present a general condition to obtain subspaces that decay uniformly in a system governed by the Lindblad master equation and use them to perform error mitigated quantum computation. The expectation values of dynamics encoded in such subspaces are unbiased estimators of noise-free expectation values. In analogy to the decoherence free subspaces which are left invariant by the action of Lindblad operators, we show that the uniformly decaying subspaces are left invariant (up to orthogonal terms) by the action of the dissipative part of the Lindblad equation. We apply our theory to a system of qubits and qudits undergoing relaxation with varying decay rates and show that such subspaces can be used to eliminate bias up to first order variations in the decay rates without requiring full knowledge of noise. Since such a bias cannot be corrected through standard symmetry verification, our method can improve error mitigation in dual-rail qubits and given partial knowledge of noise, can perform better than probabilistic error cancellation.
\end{abstract}
    \maketitle

Decoherence remains a major obstacle in performing useful quantum computation and thus keeping it from fulfilling its promises~\cite{nielsen2010quantum,pellizzari1995decoherence,chuang1995quantum,preskill2018quantum, krantz2019quantum,kjaergaard2020superconducting,breuer2002theory,clerk2010introduction,lidar2019lecture}.
While significant efforts are being made to drive down the error rates at the hardware level, continued progress is taking place in the fields of quantum error correction (QEC) and quantum error mitigation (QEM) to combat realistic noise with the aim of using the current noisy intermediate scale quantum (NISQ) hardware to its full potential. 

Powerful methods to protect quantum information were developed such as by encoding it in decoherence free subspaces (DFS) that are completely immune to noise~\cite{zanardi1997error,
PhysRevLett.79.3306,
PhysRevA.57.3276,lidar1998decoherence,wu2012time,kwiat2000experimental,lidar2014review}. Such subspaces arise as a result of symmetric noise usually resulting from collective decoherence mechanisms and therefore have limited applicability in addressing local noise affecting the NISQ hardware. In a similar spirit to DFS, symmetry verification protocols for error mitigation~\cite{PhysRevA.98.062339,
mcclean2017hybrid,
PhysRevLett.122.180501} are effective at identifying and removing specific errors making the system immune to noise that breaks the symmetry of the quantum state, but have no way of removing symmetry-preserving errors. 
In Ref.~\cite{cai2021quantum}, the author develops a sampling based method allowing for partial cancellation of errors undetectable by symmetry verification.
Probabilistic Error Cancellation (PEC) on the other hand, can completely eliminate bias, however demands full knowledge of noise that may be hard to attain~\cite{temme2017error,cai2022quantum}.

In this {article}, we show that for certain realistic local noise for which a DFS offering complete protection may not exist, we can obtain subspaces that are uniformly affected by noise. The expectation values of dynamics encoded in such \emph{Uniformly Decaying Subspaces} (UDS) are unbiased estimators of noise-free expectation values and therefore can be used for error mitigation.   
Specifically, where DFS are left invariant by the action of Lindblad operators, we show that the uniformly decaying subspaces are left invariant (up to orthogonal terms) by the action of the dissipative part of the Lindblad equation. 
We present the general conditions for construction of such subspaces and apply our theory to a system of qubits and qudits undergoing strong individual relaxation with varying decay rates. 
The varying decay rates introduce a bias in estimation which cannot be detected through standard symmetry verification. We show that such a bias can be eliminated up to first order variations in decay rates using an ensemble of runs without requiring the knowledge of individual decay rates. 
Lastly we apply our theory to implement the transverse field Ising model (TFIM) in analog and circuit-model settings using dual-rail qubits~\cite{wu2022erasure,
kubica2022erasure,
teoh2023dual,
chou2023demonstrating,
campbell2020universal,
levine2023demonstrating}.
Our method improves on the standard symmetry verification for QEM used in dual-rail qubits and given partial knowledge of noise, can perform better than probabilistic error cancellation.
We note that some earlier works have pointed out that coherences or off-diagonal parts of density matrix can decay with the same rate in $PT$ symmetric Lindbladians~\cite{prosen2012p, prosen2012generic},   with only specific  observables with additional weak symmetry experiencing a uniform decay~\cite{van2018symmetry}. Our construction ensures that the entire density matrix (off-diagonal and diagonal parts) and \textit{all} observables experience a uniform decay.

\textit{Preliminaries-}
Consider a quantum system described by a density operator $\rho$ interacting with the environment. Under the standard assumptions that a Markovian environment started to couple to the system {at some point in the past}, the dynamics is completely positive and trace preserving, described by the Lindblad equation of the form
\begin{subequations} \label{eqn:Lindblad}
\begin{align}
    \dot \rho &= \mc L \rho = \mc H \rho + \mc D \rho \;, \\
    \mc D \rho &= \sum_{i=1}^m \gamma_i \mc D[A_i] \rho \;,
\end{align}
\end{subequations}
where $\mc H \rho = -i[H,\rho]$ describes the unitarily evolving part under the Hamiltonian $H$ and $\mc D[A]\rho = A\rho A^\dagger - \{A^\dagger A, \rho \}/2$ is the dissipative part with a general Lindblad operator $A$.  
We re-write the above equation in the Liouville space or third-quantization formalism~\cite{prosen2008third,krishna2022select,li2014perturbative} as $ \kket{\dot \rho} = \hat{\mc L} \kket{\rho} \;,$ where $\hat{\mc L} = \hat{\mc H} + \hat{\mc D}$. 
{We review the Liouville formalism in Appendix~\ref{app:Liouville}.} 
The expectation value of observable $O$ is then given as
\begin{align}
\langle O(t) \rangle = \bbra{O} e^{\hat{\mc L} t}|\rho(0) \rrangle = \sum_\alpha e^{\lambda_\alpha t} \bbra{\ell_\alpha}\rho (0) \rrangle \llangle O \kket{r_\alpha} \;,
\label{eqn:<O>}
\end{align}
where $\kket{\ell_\alpha}, \kket{r_\alpha}$ are bi-orthonormalized left and right eigenvectors of $\hat{\mc{L}}$ and $\lambda_\alpha$ is the corresponding eigenvalue.

\begin{figure}
\includegraphics[width=\columnwidth]{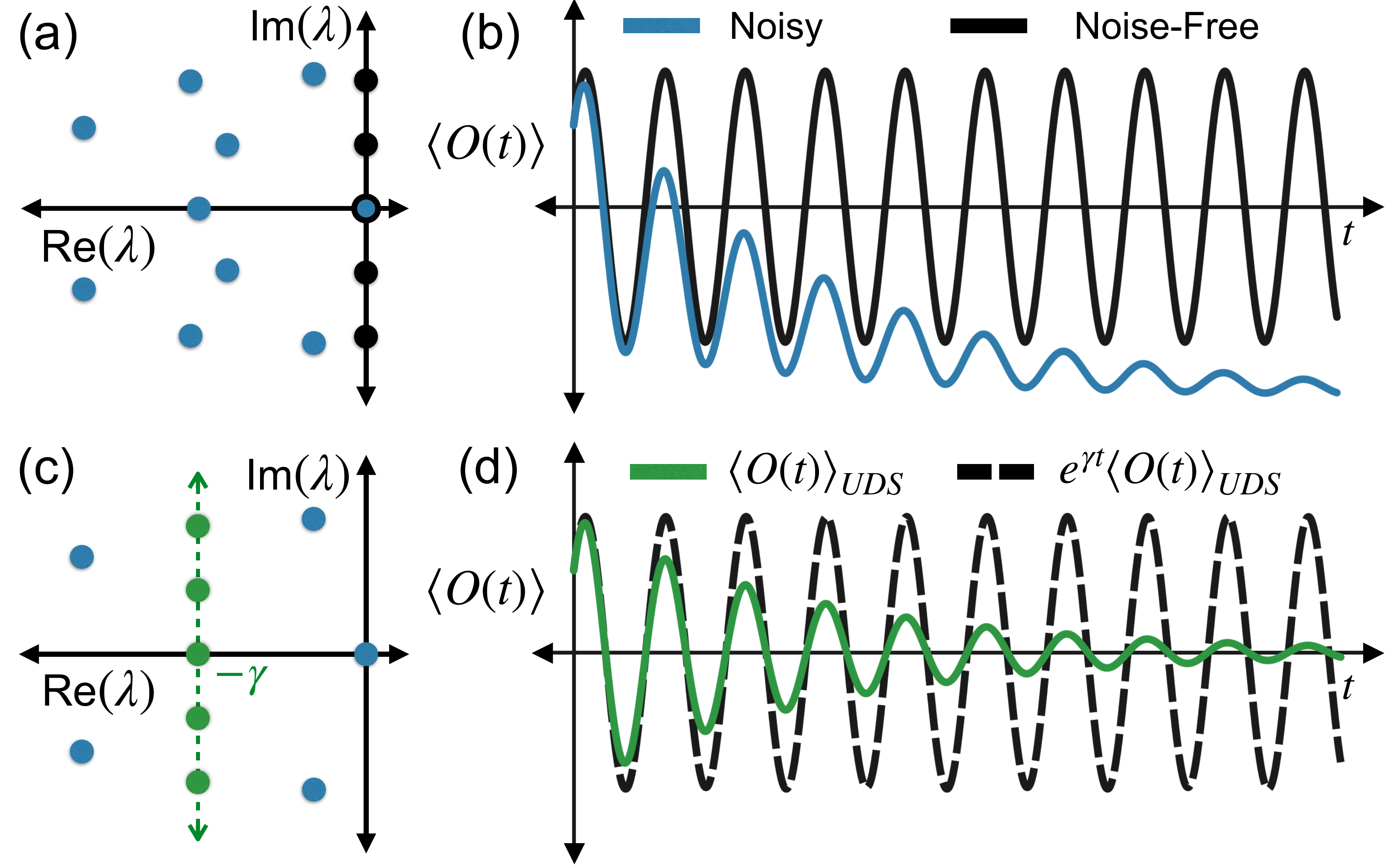}
\caption{(a) The distribution of eigenvalues of $\hat{\mc L}$ in the complex plane, with the black and blue circles indicating coherent and noisy dynamics and the corresponding (b) expectation values plotted with time. (c) The distribution of eigenvalues of $\hat{\mc L}$ for the uniformly decaying subspace marked by green circles and the corresponding (d) expectation value plotted with time.}
\label{fig:Fig1}
\end{figure}

Given the Hamiltonian $H$ such that $H\ket{E_\alpha} = E_\alpha \ket{E_\alpha},$ the eigenvalues of the Liouville operator for coherent dynamics are purely imaginary, given by $\lambda^C_{\alpha\beta} = -i (E_\alpha-E_\beta)$ and marked by black circles in Fig.~\ref{fig:Fig1}(a). The presence of noise causes the eigenvalues to be scattered in the complex plane as marked by the blue circles  forcing different parts of the Hilbert space to decay at different rates and ruining the system's coherence. As an example, one qubit experiencing relaxation with decay rate $\gamma$ already has two different decaying timescales $T_1=1/\gamma$ for diagonal and $T_2=1/(2\gamma)$ for off-diagonal parts of the density matrix. Coherence of systems is therefore preserved only for time scales $t\ll 1/\gamma$. The expectation value of an observable corresponding to the noisy and noise free cases are shown in Fig.~\ref{fig:Fig1}(b). 

{In cases with high symmetry, usually resulting from collective decoherence mechanisms, there can exist decoherence-free subspaces. 
If the Lindblad operators $A_i$ are normal and share a common set of eigenvectors $\ket{\mc P_j}$, where $A_i \ket{\mc P_j} = c_i \ket{\mc P_j} \forall i,j$, then the subspace $\mc S$ spanned by $\{\ket{\mc P_j}\}_{j=1}^N$ is protected from decoherence as $\mc D\ket{\mc P_i}\bra{\mc P_j}=0$~\cite{lidar1998decoherence}. 
The later work \cite{karasik2008criteria}  obtained a necessary and sufficient condition for the above when the $A_i$ are not necessarily normal, by noting that in addition to the previous constraint,  $\sum_i \gamma_i A_i^\dagger A_i$ should also act by a scalar on $\mc{S}$: $\sum_i \gamma_i A_i^\dagger A_i \ket{\mc P_j}= \sum_i \gamma_i |c_i|^2 \ket{\mc P_j}$. Further, if $H$ leaves $\mc{S}$ invariant, then the subspace $\mc S$ is a DFS. Then, the Lindbladian is composed of only the Hamiltonian term, and the dynamics is identical to the noise-free case in Fig.~\ref{fig:Fig1} (a),(b). }

\textit{Uniformly Decaying Subspaces-} 
Analogous to the DFS, for certain special cases if instead the entire Lindblad dissipator operator $\mc D$ leaves a set of vectors $\ket{\mc P_i}\bra{\mc P_j}$ unaffected up to an orthogonal component, then there exists a uniformly decaying subspace $\mc S$ that is spanned by $\{ \ket{\mc P_j}\}_{j=1}^N$. 
The condition is given by
\begin{align}
\mc {D}(\ket{\mc P_i} \bra{\mc P_j} \oplus \mc P^\perp) = -\gamma \ket{\mc P_i} \bra{\mc P_j} \oplus \mc P^{\prime\perp} \;,  
\label{eqn:UDS_condition}
\end{align}
$\forall i,j$ where $\mc P^\perp, \mc P^{\prime\perp}$ are in a subspace orthogonal to operators on $\mc S$. 
A sufficient condition can be obtained in terms of the Lindblad operators in a similar spirit to DFS. 
{Instead of the eigenvalue condition, we require that $A_i\ket{\psi}\in\mc{S}^\perp$ for any $\ket{\psi} \in \mc S \oplus \mc S^\perp$, where the subspace $\mc{S}^\perp$ is orthogonal to $\mc{S}$.  
Similarly to DFS, we require $\sum_i \gamma_i A_i^\dagger A_i$ to act by a scalar $-\gamma$ on $\mc{S}$: $\sum_i \gamma_i A_i^\dagger A_i \ket{\mc P_j} = -\gamma \ket{\mc P_j}$. 
We will later loosen this second restrictive condition by showing that the UDS can be made resistant to first order deviations from the scalar $-\gamma$, using the example of inhomogeneous qubit and qudit relaxation.
More general discussion is provided in Appendix~\ref{app:UDS}.}


The condition Eq.~\ref{eqn:UDS_condition} for uniformly decaying subspaces can be rewritten in the Liouville formalism as $\hat{\mc P}\hat{\mc D} = - \gamma \hat{\mc P}$, where $\hat{\mc P}=P \otimes {P^T}$ and $P = \sum_i \ket{\mc P_i}\bra{\mc P_i}$ is the projector onto $\mc{S}$. We encode the initial density matrix and observable in the subspace: $\kket{\rho(0)}=\hat{\mc P} \kket{\rho(0)}$ and $\kket{O}=\hat{\mc P} \kket{O}$. We require that {our chosen} Hamiltonian must commute with $ P$, implying $\hat{\mc P} \hat{\mc H} = \hat{\mc H} \hat{\mc P} =  \hat{\mc P} \hat{\mc H}\hat{\mc P}$. Using the conditions above, we evaluate the expectation value of the observable in the subspace after evolving for a time $t$
\begin{subequations} 
\label{eqn:UDS_expect}
\begin{align}
    \langle O(t) \rangle_{\text{UDS}} &= \bbra{O}\hat{\mc P} e^{\hat{\mc L} t} \hat{\mc P}\kket{\rho(0)}  \;, \\
    &= \sum_n \frac{t^n}{n!} \bbra{O} \hat{\mc P} (\hat{\mc H} - \gamma )^n \hat{\mc P}\kket{\rho(0)}\;,\\
    &= e^{-\gamma t} \langle O\rangle_C\;, \label{eqn:decay}
\end{align}
\end{subequations}
where $\langle O\rangle_C \equiv \bbra{O} e^{\hat{\mc H} t} \kket{\rho(0)}$ is the noise-free or coherent expectation value. We therefore see that the expectation value of an observable encoded in the subspace is an unbiased estimator of noise-free expectation value up to an overall decaying factor. The eigenvalues of the encoded dynamics are $\lambda_{\alpha\beta} = - \gamma + i(E_\alpha-E_\beta)$ shown with green circles in Fig.~\ref{fig:Fig1} (c) and are identical those for coherent dynamics translated by a common constant. The measured expectation value in the subspace is shown in Fig.~\ref{fig:Fig1} (d) with the green line which when rescaled gives back the coherent dynamics shown with the dashed black line. 

{We estimate the expectation value of the observable using $N_{\text{UDS}}$ number of shots with the estimator $\langle \hat O \rangle_{\text{UDS}}
= \sum_i\langle O\rangle_{\text{UDS},i}/{N_{\text{UDS}}}$, where $\langle O\rangle_{\text{UDS},i}$ is the $i^{\text{th}}$ measurement outcome.  To obtain the estimate with the same confidence as that of the noise-free case we require that $\text{Var}\langle \hat O\rangle_{\text{UDS}} = \text{Var}\langle \hat O\rangle_{\text{C}}$, where  $\langle \hat O\rangle_{\text{C}}= \sum_i\langle O\rangle_{C,i}/{N_{C}}$ is the estimator for noise-free case. Then, we use Eq.~\ref{eqn:decay} to obtain the sampling overhead $N_{\text{UDS}}/N_C = e^{2\gamma t}$. The elimination of bias therefore comes at the cost of requiring exponentially more measurements compared to the noise free case, which is often true for error mitigation methods~\cite{cai2022quantum}.}

{The observable $\hat{\mathcal P}\kket{O}$ or equivalently $POP$ in the UDS subspace can be written in Pauli-decomposition as $POP = \sum_{i=1}^K \alpha_i \bm \sigma_i$, where each Pauli string $\bm \sigma_i \in \{ I, X, Y, Z\}^{\otimes n}$ can be measured separately. Monte Carlo sampling can be used to achieve the sampling overhead $(\sum_{i=1}^K |\alpha_i|)^2$~\cite{cai2022quantum}.}
Alternatively, we can measure $O$ after measuring $P$, post-selecting for the case $P=1$, using an ancilla qubit, which follows the scheme behind symmetry verification~\cite{PhysRevA.98.062339, PhysRevA.98.062339,
mcclean2017hybrid,
PhysRevLett.122.180501}. 
{For example, for each dual rail qubit~\cite{kubica2022erasure, levine2023demonstrating} where ${\ket{0}_L\equiv\ket{01},\ket{1}_L \equiv \ket{10} }$, an additional ancilla qubit is required to check for decay to $\ket{00}$ state by measuring $P=I-\ket{00}\bra{00}$.} The state is measured in the computational subspace $\mc S$ with probability $\llangle  P| e^{\hat{\mc L} t}| \rho(0)\rrangle = e^{-\gamma t}$, and the post-selected expectation value is $\langle O(t)\rangle_{\text{UDS}}^{\text{Post}} = \llangle O| \hat{\mc P} e^{\hat{\mc L}t} \hat{\mc P}| \rho(0) \rrangle/\llangle  P| e^{\hat{\mc L} t}| \rho(0)\rrangle= \langle O \rangle_C$. 
The uniformly decaying part now represents the exponentially decreasing probability of finding the system in the correct subspace. 
Any errors taking the system out of {the} computational subspace are removed by post-selection. Under post-selection, the UDS conditions can be viewed as equivalent to that of symmetry verification with perfect error detection. We now extend its applicability by allowing for perturbations undetectable by symmetry verification.


\textit{Qubits/qudits under inhomogeneous relaxation-} We apply our theory to $n$ qubits subject to strong individual relaxation, described by Eq.~\ref{eqn:Lindblad} with the dissipator operators $A_i = \sigma_i^- = \ket{0}\bra{1}$ and decay rates given by $\gamma_i$.  
Such noise is a generic realistic scenario in many NISQ devices for example, in cQED superconducting qubits~\cite{PhysRevLett.120.260504, burnett2019decoherence} and photonic qubits  where the dominant error is photon loss~\cite{teoh2023dual,
chou2023demonstrating}. 
In transmon qubits, such a noise bias can be engineered by operating them in resonance forming the dual-rail qubits with relaxation as the dominant error~\cite{kubica2022erasure,
campbell2020universal,levine2023demonstrating}. Moreover, dynamical decoupling can be used to further suppress dephasing leading to an even greater increase in the bias~\cite{viola1999dynamical,lidar2014review,de2010universal,
biercuk2009optimized}. The decay rates for different qubits are different and even fluctuate over time~\cite{carroll2022dynamics}.

We assume that we do not know the individual decay rates $\gamma_i$ and only have the knowledge of the average decay rate $\bar{\gamma} = \frac{1}{N} \sum_{i=1}^N \gamma_i$. We parameterize $\gamma_i = \bar{\gamma} + \Delta \gamma_i$ and write Eq.~\ref{eqn:Lindblad} as $\mc L \rho = \mc L_0 \rho + \mc L_1 \rho \;,$ where
\begin{align}
\mc L_0 \rho &= \mc H \rho + \bar \gamma \sum_i \mc D[\sigma_i^-]\rho \;,\\
\mc L_1 \rho &= \sum_i \Delta \gamma_i \mc D[\sigma_i^-]\rho \;.
\end{align}
We provide the solution to the above equations by solving for the $\mc L_0 \rho$ term exactly and treat the second term $\mc L_1 \rho$ perturbatively. We require that the decay timescales of qubits are close to one another or $|\Delta \gamma|_{\max} \ll \bar \gamma ,\| H \|$. 
We have $\|  \hat{\mc L}_1 \| = |\Delta \gamma|_{\max} = \max_i |\Delta \gamma_i|$. The condition above poses no restrictions on the average decay time scales, which in principle can be large. 

We consider the basis $\ket{\mc P_{\bm\alpha}} = \ket{\alpha_1 \dots \alpha_n}$, where $\alpha_i \in \{0,1\}$ denotes the state of the $i\text{th}$ qubit. 
We define our computational subspace  $\mc S_k$ as the space spanned by all $\ket{\mc P_{\bm\alpha}}$ with $\sum_{i=1}^n \alpha_i = k$. 
In other words, each vector in this constant Hamming weight subspace has $k$ qubits in state $\ket{1}$ and $n-k$ qubits in state $\ket{0}$. 
Therefore the dimension of the computational subspace is given by $N = \text{dim}(\mc S_k) = {n \choose k}$. 
We see that the dissipation operator of the unperturbed term $\mc D_0 \equiv \bar\gamma \sum_i \mc D[\sigma_i^-]$ satisfies Eq.~\ref{eqn:UDS_condition} with $\gamma = k \bar\gamma$ where $\mc S^\perp = \oplus_{i=0}^{k-1} \mc S_i$. 
Following the rest of the derivation, we see from Eq.~\ref{eqn:UDS_expect} that $\langle O(t) \rangle_{0} = \bbra{O}\hat{\mc P} e^{\hat{\mc L_0} t} \hat{\mc P}\kket{\rho(0)} = e^{-k\bar\gamma t} \langle O\rangle_C$. Therefore, given $n$ qubits decaying with the same rate, we can obtain an unbiased estimation of expectation values by encoding the dynamics in $\mc S_k$. It is worthwhile to note that even for this apparently symmetric unperturbed term $\mc D_0$, there exists only a trivial DFS with one ket $\ket{0\dots 0}$, which cannot be used for computation.

We now consider the second term $\mc{L}_1$ perturbatively, calculating the bias in the estimation of expectation value of the observable $\langle O \rangle$ as $\Delta \langle O \rangle= \langle O \rangle_{\text{UDS}} - e^{-k\bar\gamma t} \langle O \rangle_C$. We have
\begin{align}
 \Delta \langle O \rangle &= \llangle O| e^{\hat{\mc L}_0 t}  \int_{0}^t dt^\prime \hat{\mc L}_1^I(t^\prime) \kket{\rho(0)} + \mc O(|\Delta \gamma|_{\max}^2)\;,
\end{align} 
where $\hat{\mc L}_{1}^I(t) \equiv e^{-\hat{\mc L}_0 t} \hat{\mc L}_1 e^{\hat{\mc L}_0 t}$. 
We see that the variation of decay rates introduce{s} additional bias that cannot be detected just by symmetry-verification.
{We define a unitary translation operator $ T$ that shifts the qubits by one position, acting on the basis as $T \ket{\alpha_1 \alpha_2 \dots \alpha_{n-1}\alpha_n} = \ket{\alpha_n\alpha_1 \dots \alpha_{n-2}\alpha_{n-1}}$, which is also equivalent to qubit relabeling. We use this shifted basis for computation, i.e. rotate the initial state, Hamiltonian and observable by $T$,} and see that it leaves the unperturbed term invariant and only affects the perturbed term as 
\begin{subequations} 
\label{eqn:translate}
\begin{align}
    \mc L_1 ( T \rho  T^\dagger) &=  T \sum_i \Delta \gamma_{i-1} \mc D[\sigma_i^-](\rho)  T^\dagger \;.
\end{align}
\end{subequations}

If we consider an ensemble of $n$ evolutions with encodings shifted by one each corresponding to $ T^j$ where $j=0,\dots, n-1$ and take the average of expectation values, the first order term is proportional to $\sum_{j=1}^n \Delta \gamma_{i-j} = \sum_{k=1}^n \Delta \gamma_k =0$ as $\gamma_i = \bar\gamma + \Delta \gamma_i$. The bias $\Delta\langle O \rangle$ after taking the average over $n$ evolutions is given by
\begin{align}
  {e^{k\bar\gamma t} \llangle O \rrangle_{\text{UDS}} -  \langle O \rangle_C  = \mc O\bigg({k^2 |\Delta \gamma|_{\max}^2 t^2 \| O\|}\bigg) \;,}
 \label{eqn:UDS_avg}
\end{align}
where $\llangle O \rrangle_{\text{UDS}} =  \frac{1}{n}\sum_{j=1}^n \langle O \rangle_{j,\text{UDS}}$. We therefore see that the bias is only second order in deviation of decay rates.  The cancellation works similarly for the post selected case. Importantly, since the bias only depends on the deviation of the decay rates, and not on the magnitude of their average, it is possible to lower the bias arbitrarily by  worsening the decay rates of all qubits until they align with the fastest decaying qubit, for example by tuning operating frequencies of the qubits. This comes at the expense of requiring more shots as the average decay rate increases.  It is also noteworthy to point out that the knowledge of individual decay rates is not required to achieve this cancellation.

Furthermore, we also consider the more general case of amplitude damping channels where $A_i = a_i$ is the the bosonic annihilation operator. This noise model   is especially important for  photonic $d$-level qudits as photon loss is the dominant source of noise \cite{chuang1997bosonic, gardiner2004quantum}. 
Our calculations generalize in a straightforward manner, with the details given in Appendix~\ref{app:UDS}. 

\begin{figure}[t]
\includegraphics[width=\columnwidth]{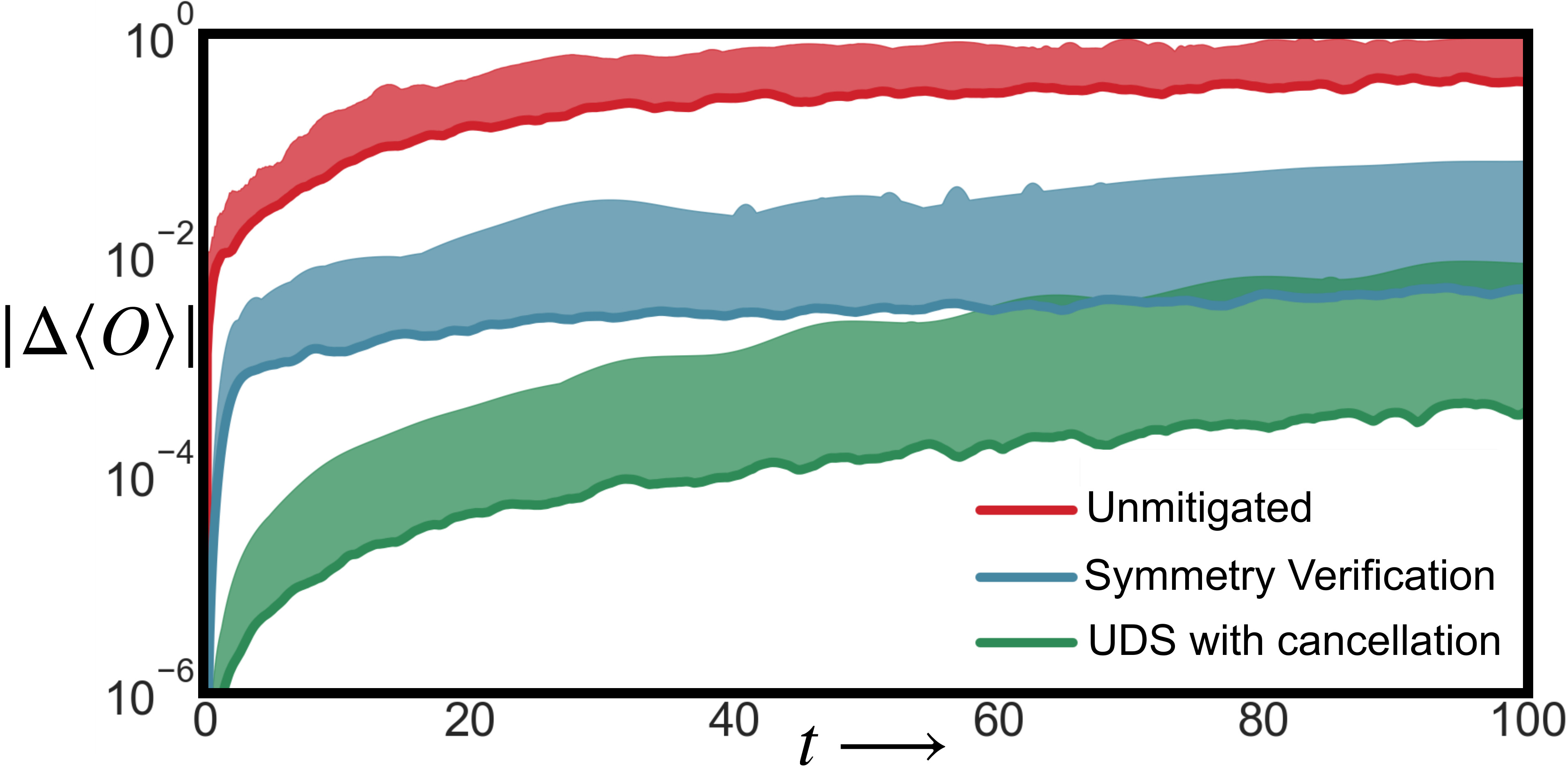}
    \caption{We consider six physical qubits experiencing individual relaxation and encode a three qubit TFIM Hamiltonian in the dual-rail logical subspace. We chose initial state to be $\ket{0}_L^{\otimes 3}$ and measure the $Z$ component of the first spin. The system is evolved using the Lindblad master equation and the modulus of the bias averaged over $100$ random samples is shown by solid lines, with the boundary of shaded area indicating the worst case bias.}
    \label{fig:Fig2}
\end{figure}

\textit{Examples-} We now illustrate the QEM method by simulating the random transverse field Ising model (TFIM) in analog and circuit-model settings. Given $n$ physical qubits, we choose $k=n/2$ to maximize the encoding efficiency, however we restrict our computational subspace to dual-rail qubits $\{\ket{0}_L\equiv\ket{01},\ket{1}_L \equiv \ket{10}\}$ in order to efficiently implement the Ising Hamiltonian. The number of logical qubits is then limited to $n_L=n/2$, which is much less than $
\log_2{n\choose {n/2}}$, but importantly this preserves the tensor product structure and allows the TFIM Hamiltonian to be realized using only two body local interactions.   The TFIM Hamiltonian is given by 
\begin{align}
    H = \frac{1}{2}\sum_{ij=1}^{n_L} J_{ij} Z_i^LZ_j^L + \frac{1}{2}\sum_{i=1}^{n_L} h^x_i X^L_i\;,
\end{align}
where $Z^L_i \equiv Z_{2i}$ and $X^L_i \equiv (X_{2i} X_{2i+1} + Y_{2i} Y_{2i+1})/2$, are the logical operations on $i^{\text{th}}$ qubit where $i = 0,\dots,n_L-1$.
 We can see that it preserves the dual-rail subspace and therefore commutes with the projector $P$ to that subspace. We take $100$ random instances  with parameters $J_{ij}, h^x_i$ chosen from a uniform distribution over $[-1,1]$ and decay rates $\gamma_i$ from a Gaussian distribution with mean $10^{-2}$ and standard deviation $10^{-3}$.

\textit{Analog-} We take six physical qubits that form three dual-rail logical qubits to simulate the TFIM Hamiltonian by solving the Lindblad equation.  We choose the initial state to be $\ket{0}_L^{\otimes 3}$ and measure the $Z$ component of the first spin. For the UDS and dual-rail method, we choose the observable {$O =P Z^L_0 P$, where the projector $P={\prod_{i=0}^2} P_i$, such that $P_i=\frac{I-Z_{2i}Z_{2i+1}}{2}$ takes values $\{1,0\}$ dependent on whether the $i$th logical qubit is in the subspace. The observable is encoded in the computational subspace, and since it is a tensor product of Paulis, it can thus be measured efficiently. Alternatively since $Z_0^L$ individually commutes with all $Z_i$, one can use post-selection by measuring all $Z_i$ simultaneously, keeping only the measured value $Z^L_0$ when each $P_i=1$.}
The modulus of the bias $\lvert \Delta \langle O \rangle\rvert$ is plotted in Fig.~\ref{fig:Fig2} on a log scale. The solid lines (bottom of the envelope) indicate the average bias of the $100$ random samples and the boundary of the shaded region indicates the maximum bias. For reference, we plot the unmitigated bias which is calculated without any postselection.
The dual-rail qubits use post-selection for symmetry verification eliminating any symmetry breaking errors. However due to the varying decay rates there is additional undetected error within the computational subspace.
For the UDS method, we consider six separate evolutions with the encodings shifted by one each time as given in Eq.~\ref{eqn:translate} and calculate their average as given by Eq.~\ref{eqn:UDS_avg}. This averaging procedure allows for the elimination of errors that are undetected by symmetry verification up to the first order and dramatically reduces the bias.

\begin{figure}[t!]
\includegraphics[width=\columnwidth]{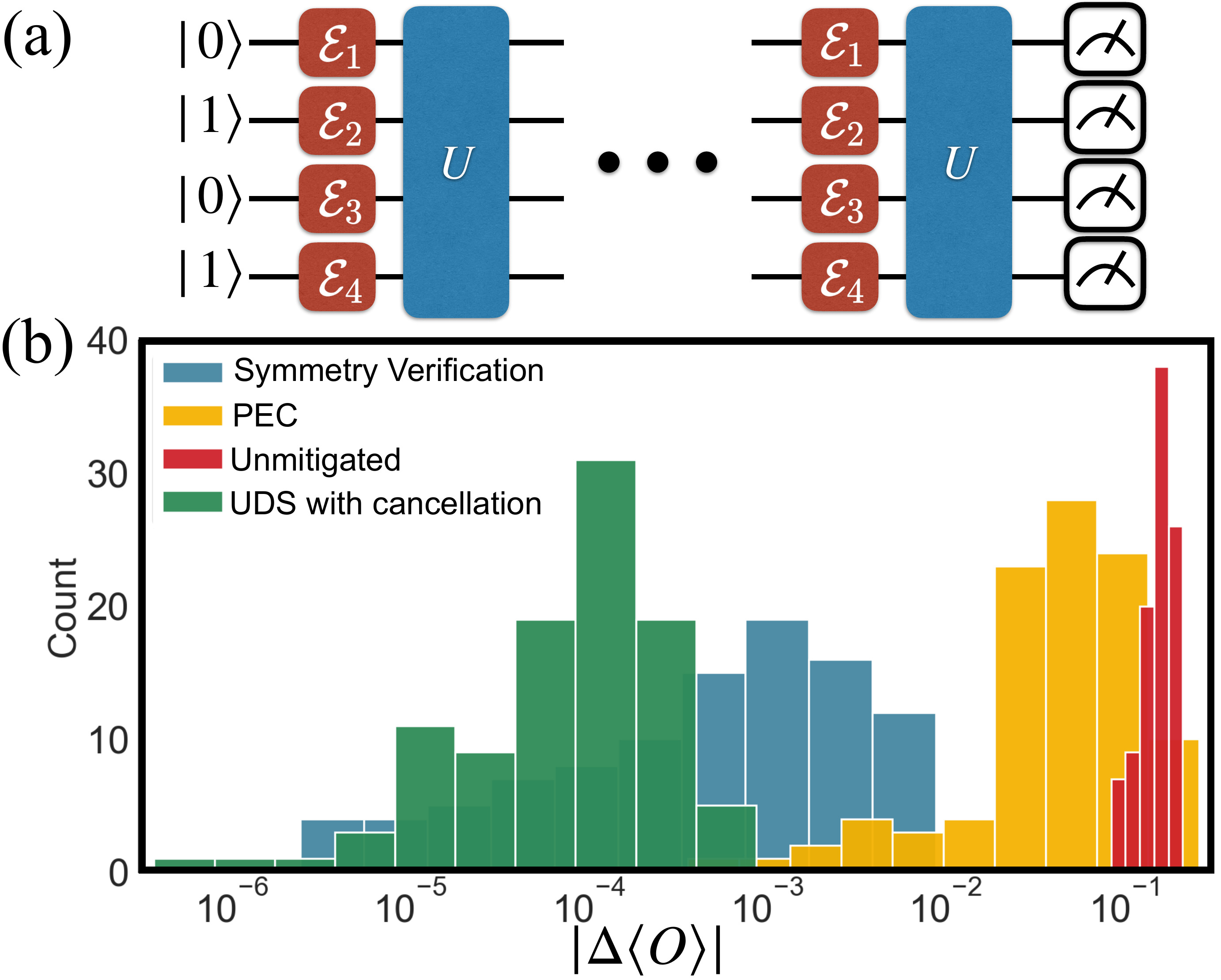}
    \caption{We consider four physical qubits experiencing individual relaxation and encode a two qubit TFIM Hamiltonian in the dual-rail subspace. The Hamiltonian is implemented using $20$ Trotter steps, where each unitary takes time $1/20$ and is implemented as a product of one and two qubit gates. The amplitude damping channel acts on the $i^{\text{th}}$ qubit with damping rate $\gamma_i$. We choose the initial state $\ket{0}^{\otimes 2}_L$ and measure the qubits in computational basis.  We allocate the same number of shots $ N_{\text{shots}} = 10^{7}$ to each method and plot the histograms of the absolute bias of $100$ random samples.}
    \label{fig:Fig3}
\end{figure}

\textit{Circuit model-} Where the previous example was focused on exact calculations of expectation values, we now show the applicability of our method on digital quantum computers with finite shots.  We trotterize the evolution into $ N_T$ Trotter steps, applying $ N_T$ unitaries decomposed as one and two qubit gates with gate time $t/ N_T$.  
Rather than evolving with respect to the Lindblad master equation Eq.~\ref{eqn:Lindblad}, the dissipation is now modeled by the discrete amplitude damping channel $\mc{E}_i$ acting on the $i^{th}$ qubit with damping rate $\gamma_i$. 
We take four physical qubits, therefore encoding two dual-rail logical qubits, and choose an illustrative case fixing $ N_T=20$ Trotter steps and time $t=1$. 
We allocate the same number of shots $ N_{\text{shots}} = 10^{7}$~\footnote{The results remain qualitatively the same for much less number of shots, we chose $10^7$ to clearly show the plots on the log scale.} to each method and measure the $Z$ component of the first spin same as above. We plot the histograms of the absolute bias of $100$ random samples on a log scale in Fig.~\ref{fig:Fig3}. 
We see that given the same number of shots, the UDS averaging method is able to perform better than dual-rail encoding alone by being able to eliminate first order bias undetected through symmetry verification. Since we also assume that only the average decay rates are known, probabilistic error cancellation can also only eliminate bias up to zeroth order. {We verify that the UDS results generalize to the discrete channel setting in Appendix~\ref{sec:circuit}.}

{For completeness, we remark that in the example above, the observable commutes with individual $Z_i$, allowing for the simultaneous measurement of $O$ and $Z_i$. This need not be the case in general. For example, if we choose $O=P X_0^L P$, since $X_0^L$ does not commute with individual $Z_i$, we therefore cannot measure $O$ and $Z_i$ simultaneously. One option is to use post-selection, i.e. use an ancilla qubit to first measure each $P_i$ and only proceed to measure $X_0^L$ when $P_i=1$. Note that this requires additional number $n_L$ of ancilla qubits. This can be done at the circuit level by requiring $2 n_L$ CNOT gates to measure $P$~\cite{PhysRevA.98.062339} or done at pulse level in the hardware directly~\cite{kubica2022erasure, levine2023demonstrating}. Alternatively, since $O$ is a sum of Paulis, explicitly $O = \frac{1}{4}(X_0X_1 + Y_0Y_1 - X_0X_1 Z_2 Z_3 - Y_0 Y_1 Z_2 Z_3)$, it can be measured efficiently by measuring individual terms or using Monte Carlo sampling~\cite{cai2022quantum}, without any additional qubit overhead.}

{Although the example above discusses the implementation of the TFIM Hamiltonian specifically, we note that the UDS method and the first order cancellation is applicable for any arbitrary quantum circuit as the dual rail qubits can implement universal quantum computation~\cite{lidar2014review}. The dual rail encoding requires $2 n_L$ physical qubits to encode $n_L$ logical qubits. We can realize an arbitrary SU(2) rotation on the logical qubit $i$ by using the Euler rotations $e^{i\gamma X_i^L} e^{i\beta Z_i^L} e^{i\alpha X_i^L}$, which together with logical control Z gate $CZ^L_{i,j}=CZ_{2i+1,2j+1}$ form a universal gate set. This requires  native XY gate and control Z operations, readily available in many platforms. Additionally, to perform the first order cancellation, we require the physical qubits to be connected in a ring topology to implement evolution with respect to the shifted basis. We can see this by considering the shifted Hamiltonian $THT^\dagger$, which requires XY connectivity between the first and last physical qubit. For measurements as discussed above, no ancilla qubit is required if we use Monte Carlo sampling, and $n_L$ ancilla qubits are needed if we use post-selection to measure each $P_i$ first.} 

\textit{Discussion-} 
{We presented conditions to obtain subspaces that decay uniformly and used them to perform error mitigated quantum computation. Where DFS are left invariant by the action of Lindblad operators, we show that the uniformly decaying subspaces are left invariant (up to orthogonal terms) by the action of the dissipative part of the Lindblad equation. For localized noise where a DFS offering complete protection may not exist, we show that we can still encode quantum information into UDS which can be used for error mitigation.  We applied our theory to a system of qubits undergoing strong individual relaxation with varying decay rates and showed that the first order bias can be eliminated using an ensemble of runs without requiring the knowledge of individual decay rates. The subspaces therefore can be used to obtain unbiased estimators of the true expectation value up to second order variations in decay rates. We demonstrated our method by implementing the transverse field Ising model in analog and digital settings obtaining lower bias estimates as compared to symmetry verification and probabilistic error cancellation methods.} 
Lastly, our framework can be extended to other noise models, such as correlated decays with $A_{ij} = \sigma_i^- \otimes \sigma_j^-$, which similarly give a UDS with cancellation of first-order errors, discussed in Appendix~\ref{app:UDS}. However, the cancellation becomes less straightforward with inefficient scaling. {Finally, the UDS method may also be naturally suited for Fermionic Hamiltonian simulation problems that naturally conserve particle number~\cite{setia2018bravyi,bravyi2002fermionic}}. The techniques introduced open up new possibilities in achieving lower biases than previously possible with symmetry verification for error mitigation,  without requiring complete knowledge of noise. Since our method is a passive encoding scheme, it can be used as a base encoding  with other QEM or QEC methods.

\textit{Acknowledgements-} This material is based upon work supported by the U.S. Department of Energy, Office of Science,
National Quantum Information Science Research Centers, Superconducting Quantum Materials
and Systems Center (SQMS) under the contract No. DE-AC02-07CH11359 (NASA-DOE IAA SAA2-403602).
N.S. and D.V. acknowledge support under NASA Academic Mission Services under contract No. NNA16BD14C. All authors acknowledge helpful discussions with Jeffrey Marshall, Stuart Hadfield, Tanay Roy, Namit Anand, Eliot Kapit, Jacob Biamonte and the NASA QuAIL group.


%

\onecolumngrid
\appendix 

\section{Liouville Space Representation}
\label{app:Liouville}
Let $\mc{W}$ be a Hilbert space of pure states. For any subspace $\mc{V}\subseteq\mc{W}$, let $\omc{V} = \textnormal{End}(V)$ denote the space of linear operators $\mc{V}\rightarrow\mc{V}$. 

For $\rho\in\omc{W}$, the Lindblad master equation is given as $\dot\rho = \mc L \rho = \mc H \rho + \mc D \rho \;,$ such that
\begin{align}
\mc H \rho &= -i[H,\rho] \;, \\
\mc D\rho &= \sum_i \gamma_i \mc D[
A_i] \rho \;,
\end{align}
where $\mc D[A] \rho = A\rho A^\dagger - \frac{1}{2}\{A^\dagger A,\rho\}$\;.
In many scenarios, analyzing the equation and its evolution becomes easier in the vectorized or Liouville space representation also known as the third-quantization. The rows of $N \times N$ density matrix $\rho$ can be stacked on top of one another to obtain its vectorized form $\kket{\rho}$ with dimension $N^2 \times 1$, identifying $\omc{V}$ with $V\otimes V^*$. As an example the $2 \times 2$ density matrix transforms as $\rho = \begin{bmatrix}
\rho_{00} &\rho_{01}\\
\rho_{10} & \rho_{11}
\end{bmatrix} \iff \kket{\rho} = \begin{bmatrix} \rho_{00} &\rho_{01} &
\rho_{10} & \rho_{11}\end{bmatrix}^T$. 
Given a superoperator $\mc{F}$ on ${\omc{V}}$, we denote the corresponding operator in the Liouville space representation as $\hat{\mc{F}}$. 
Using the operator transformation rule $A \rho B \iff A \otimes B^T \kket{\rho}$, we can write the above Lindblad master equation in the Liouville space $\kket{\dot \rho} = \hat{\mc L} \kket{\rho}\;,$ where 
\begin{align}
    \hat{\mc L} &= \hat{\mc H} + \hat{\mc D} = -i (H\otimes I - I \otimes H^T) + \sum_i \gamma_i \bigg(A_i\otimes A_i^* - \frac{1}{2} (A_i^\dagger A_i \otimes I + I \otimes A_i^T A_i^*)\bigg)\;. 
\end{align}
We can now diagonalize the superoperator
\begin{align}
\hat{\mc L}\kket{r_\alpha} &= \lambda_\alpha \kket{r_\alpha} \;, & \hat{\mc L}^\dagger\kket{l_\alpha} &= \lambda_\alpha^* \kket{l_\alpha} \;,
\end{align}
where $\kket{r_\alpha}, \kket{l_\alpha}$ are the corresponding right and left eigenvectors and $\lambda_\alpha$ is the eigenvalue. Bi-orthonormality can be imposed such that $\bbra{l_\alpha}r_\beta\rrangle = \delta_{\alpha \beta}$, where the inner product in the Liouville space is equivalent to the one defined for operators $\llangle A|B \rrangle = \tr(A^\dagger B)$. The evolution of the density matrix is then given as 
\begin{align}
\kket{\rho(t)} = \sum_\alpha e^{\lambda_\alpha t} \bbra{l_\alpha}\rho (0) \rrangle \kket{r_\alpha} \;{.}
\end{align}
The expectation value of an observable can be calculated as 
\begin{align}
\langle O \rangle = \bbra{O} \rho(t) \rrangle = \sum_\alpha e^{\lambda_\alpha t} \bbra{l_\alpha}\rho (0) \rrangle \llangle O \kket{r_\alpha} \;.
\end{align}
We can also write the Lindblad-like form for the corresponding Liouville representation of $\hat{\mc L}^\dagger$ given by 
\begin{align}
\mc L^\dagger \rho = + i[H,\rho] + \sum_i \gamma_i \mc D^\dagger [A_i] \rho \;,
\end{align}
where $\mc{D}^\dagger [A] \rho = A^\dagger \rho A - \frac{1}{2} \{ A^\dagger A, \rho\} \;.$ 

\section{Uniformly Decaying Subspaces}\label{app:UDS}
\subsection{General Derivation}\label{sec:general derivation}
We present the general conditions to achieve Uniformly Decaying Subspaces (UDS). 
As above, we let $\mc{W}$ be the entire (pure) state space and let $\mc{V}$ be a chosen subspace. 
The expectation value of an observable $O$ is given as 
\begin{align}
    \langle O \rangle = \bbra{O} e^{\hat{\mc L} t} \kket{\rho(0)}\;,
    \label{eqn:Expect}
\end{align}
where ${\hat{\mc L}} = \hat{\mc H} + \hat{\mc D}$ is a general vectorized Lindblad operator. 
Let $\mc{S}$ be a subspace of the state space $\mc{V}$, with space of operators $\omc{S} = \textnormal{End}(\mc{S})$ and projector $P$ onto $\mc{S}$. 
We write ${\mc P}$ to denote the corresponding projector onto $\omc{S}$, with
\begin{align}
    \mc{P}(\rho) &= P\rho P
    \\ \hmc{P}\kket{\rho} &= P\otimes {P^T} \kket{\rho}
\end{align}
in the operator and Liouville space representations respectively. 
We will use the notation $\omc{S}^\perp$ to denote the orthogonal complement of $\omc{S}$ in $\omc{V}$ (not the full orthogonal complement in $\omc{W}$). 
We obtain a uniformly decaying subspace if 
\begin{align}\label{uds_assump_unified}
    \hmc{P}\hmc{D} = -\gamma \hmc{P}
\end{align}
as operators on $\omc{V}$. 
In other words, $\hmc{P}\hmc{D}$ is a diagonalizable operator on $\omc{V}$, with eigenvalue $-\gamma$ on $\omc{S}$ and eigenvalue $0$ on $\omc{S}^\perp$. 
We will often find it more intuitive to decompose this into two equalities: 
\begin{align}
    \hmc{P}\hat{\mc D} \hat{\mc P} &= - \gamma \hat{\mc P} \;, \label{uds_assump_scalar} \\
    \hat{\mc P} \hat{\mc D} &= \hat{\mc P} \hat{\mc D} \hat{\mc P}\;,
    \label{eqn:D}
\end{align} 
as operators on $\omc{V}$. 
The first tells us that $\hmc{P}\hat{\mc D} \hat{\mc P}$ has eigenvalue $-\gamma$ on $\omc{S}$. 
The second tells us that $\hmc{D}$ maps $\omc{S}^\perp$ to $\omc{S}^\perp$. 
Viewing $\hmc{D}$ as a matrix, we have
\begin{equation}
    \hmc{D} = \begin{bmatrix}
        -\gamma I & 0\\ * & *
    \end{bmatrix}.
\end{equation}
In more mathematical language, we would say that $\hmc{D}$ leaves the spaces $\omc{V}$ and $\omc{S}^\perp$ invariant, and therefore that $\hmc{P}\hmc{D}\hmc{P}$ is an operator on the quotient space $\omc{V}/\omc{S}^\perp$ (which is isomorphic to $\omc{S}$ as a vector space). 
Eq.~\eqref{uds_assump_scalar} is the assumption that this operator on the quotient space is a scalar. 


We construct a Hamiltonian that commutes with $\mc P$, implying
\begin{align}
        \hat{\mc H} \hat{\mc P} = \hat{\mc P} \hat{\mc H} = \hat{\mc P} \hat{\mc H} \hat{\mc P} \;.
        \label{eqn:H}
\end{align}
We encode the initial state and observable in the subspace $\mc{S}$, taking
\begin{align}
    \kket{\rho(0)} &= \hat{\mc P} \kket{\rho(0)} \;, & \kket{O} &= \hat{\mc P} \kket{O} \;. 
    \label{eqn:Encode}
\end{align}
Using Eq.~\ref{eqn:Encode}, we rewrite the expectation value
\begin{align}\label{eqn:uds expectation}
    \langle O \rangle_{\text{UDS}} = \bbra{O} \hat{\mc P} e^{{\hat{\mc L}} t} \hat{\mc P}\kket{\rho(0)} = \sum_n \frac{t^n}{n!} \bbra{O} \hat{\mc P} (\hat{\mc H} + \hat{\mc D})^n \hat{\mc P}\kket{\rho(0)} = \bbra{O}  e^{ \hat{\mc P}{\hat{\mc L}} \hat{\mc P} t}\kket{\rho(0)} = e^{-\gamma t} \langle O \rangle_C \;,
\end{align}
where $\langle O \rangle_C \equiv \bbra{O}   e^{ \hat{\mc P} \hat{\mc H} \hat{\mc P} t}\kket{\rho(0)}$ is the expectation value of the observable in the coherent case without any dissipation. We used Eq.~\ref{eqn:D} and Eq.~\ref{eqn:H} to obtain the second last equality. We can see that the decay part factors out and the effective evolution inside the uniformly decaying subspace is coherent up to the overall scaling factor.


\subsection{Sufficient conditions using Lindblad equation}\label{sec:explicit}
We now consider the 
Lindblad equation
\begin{align}
\dot \rho = \mc L \rho = \mc H \rho + \sum_{\beta\in\mc{B}} \gamma_\beta \mc D[A_\beta] \rho \;,\label{x_assump_lindblad}
\end{align}
where $\mc{B}$ is some index set and each $A_\beta$ is an operator on the state space $\mc{W}$. 
We will give sufficient conditions on the operators $A_\beta$ to obtain a UDS as above. 
Let $\mc{A}$ be the associative algebra generated by the $A_\beta$ and $A_\beta^\dagger A_\beta$ (note we do not generally have $A_\beta^\dagger\in\mc{A}$). 
Locate a pair of nested subspaces $\mc{S}^\perp\subseteq \mc{V}$ of $\mc{W}$ that are invariant subspaces for the algebra $\mc{A}$, 
and let $\mc{S}$ (the computational subspace) be the orthogonal complement to $\mc{S}^\perp$ in $\mc{V}$, with projector $\mc{P}$. 
We in fact add the stronger assumption that $A_\beta$ maps $\mc{V}$ into $\mc{S}^\perp$. 
Graphically, we summarize these conditions as
\begin{equation}\label{eq:invariant}
    \mc{V}\xrightarrow{A_\beta^\dagger A_\beta} \mc{V},\,\hspace{1 in}
    \mc{S}^\perp\xrightarrow{A_\beta^\dagger A_\beta} \mc{S}^\perp,\, \hspace{1 in}
    \mc{V}\xrightarrow{A_\beta} \mc{S}^\perp \xrightarrow{A_\beta} \mc{S}^\perp.
\end{equation}
Equivalently, for all $\ket{\psi}\in\mc{V}$ and all indices $\beta$, we have: 
\begin{align}
    &P A_\beta\ket{\psi} = 0, \label{x_assump_0}
    \\ & P A_\beta^\dagger A_\beta \ket{\psi} = P A_\beta^\dagger A_\beta P\ket{\psi}.\label{x_assump_pd}
\end{align}
We also assume that for some scalar $c>0$, 
\begin{equation}
    P\left(\sum_\beta A_\beta^\dagger A_\beta \right)P = cP.\label{x_assump_scalar}
\end{equation}

Given the above assumptions, we let $\olg$ be the average of the coefficients $\gamma_\beta$, and define 
\begin{equation}\label{eq:def D}
    {\mc{D}_0} = \olg \sum_{\beta\in\mc{B}} {\mc{D}}[A_\beta]
\end{equation}
(with corresponding Liouville space representation ${\hmc{D}_0} = \olg \sum_{\beta\in\mc{B}} {\hmc{D}}[A_\beta]$). 
We will consider $\hmc{L}_0 = \hmc{H} + {\hmc{D}_0}$, where 
\begin{equation}
    \mc{H}\rho = -i[H,\rho],
\end{equation}
corresponding to a Hamiltonian $H$ on $\mc{V}$ that commutes with $\mc{P}$. 
Writing 
\begin{equation}
    \Delta\gamma_\beta = \gamma_\beta-\olg, 
\end{equation}
we have $\hmc{L} = \hmc{L}_0 + \hmc{L}_1$, where $\hmc{L}_1 = \sum_\beta \Delta\gamma_\beta {\hmc{D}}[A_\beta]$. 

It is straightforward to see that the operator ${\hmc{D}_0}$ satisfies the condition \eqref{uds_assump_unified} discussed above; we give the calculation explicitly here. 
Note that by \eqref{x_assump_0} and \eqref{x_assump_pd}, 
\begin{align}
    \mc{P}{\mc{D}}[A_\beta](\rho) &= P A_\beta\rho A_\beta^\dagger P -\frac{1}{2}P\left(A_\beta^\dagger A_\beta \rho + \rho A_\beta^\dagger A_\beta\right)P
    \\&= -\frac{1}{2}P\left(A_\beta^\dagger A_\beta P\rho P + P\rho P A_\beta^\dagger A_\beta\right)P.
    \\&= \mc{P} {\mc{D}}[A_\beta] \mc{P}(\rho). 
\end{align}
Summing the penultimate line over all $\beta$ and applying \eqref{x_assump_scalar}, we obtain 
\begin{align}
    \mc{P}{\mc{D}_0}(\rho) 
    &= {-\frac{1}{2}\olg\left( P\left(\sum_\beta A_\beta^\dagger A_\beta\right) P\rho P +  P\rho P \left(\sum_\beta A_\beta^\dagger A_\beta\right) P\right)}
    \\&= {-\frac{1}{2} c \olg ( P\rho P +  P\rho P)}
    \\&= -c\olg \mc{P}(\rho).
\end{align}
This is \eqref{uds_assump_unified} with $\gamma = c\olg$. 
Then by the calculations of Appendix~\ref{sec:general derivation}, we conclude that \eqref{eqn:uds expectation} holds for $\hmc{L}_0 = \hmc{H} + {\hmc{D}_0}$ as above. 

After considering some examples, we will discuss the effect of the perturbation term $\hmc{L}_1$. 
We note that the fact that $\hmc{P}{\hmc{D}}[A_\beta] = \hmc{P}{\hmc{D}}[A_\beta]\hmc{P}$ implies that 
\begin{equation}\label{eq:PL1}
    \hmc{P}\hmc{L}_1 = \hmc{P}\hmc{L}_1 \hmc{P}.
\end{equation}

\subsection{Main example: amplitude damping channel}\label{sec:main system}
We now consider the example of the bosonic amplitude damping channel \cite{chuang1997bosonic}, generalizing the qubit case considered in the main text. 
Let $\mc{W}$ be the state space of $n$ qudits, each of dimension $d$.  
Let $a = \sum_{m\geq 0} \sqrt{m} \ketbra{m-1}{m}$ be the bosonic annihilation operator, with $a_i$ denoting the annihilation operator on qudit $i$. 
We consider the Lindblad equation \eqref{x_assump_lindblad} with $\mc{B} = \{1, \dots, n\}$ and $A_i = a_i$: 
\begin{equation}\label{lindblad damping}
    \mc{L} = \mc{L}_0 + \mc{L}_1, \,\, \mc{L}_1 = \sum_i \Delta\gamma_i {\mc{D}}[A_i],
\end{equation}
where all notation is as above. 
(Note that $d=2$ gives the qubit case of the main text, with $a_i = \sigma_i^-$.) 

Let the \emph{degree} of a Fock basis vector $\ket{\alpha}=\ket{\alpha_1\cdots \alpha_n}$ be $\deg \ket{\alpha} = \sum_i \alpha_i$, 
and define $\mc{S}_k$ to be the space spanned by all Fock basis vectors of degree $k$. 
We fix a degree $k$ and consider 
\begin{equation}
    \mc{V} = \bigoplus_{i=0}^k \mc{S}_i, \,\, \mc{S}=\mc{S}_k, \,\, \mc{S}^\perp = \bigoplus_{i=0}^{k-1} \mc{S}_i. 
\end{equation}
It is easy to see that these operators satisfy the assumptions of the previous section. 
In particular, \eqref{eq:invariant} is immediate from the fact that $a_i$ is degree-lowering and $a_i^\dagger a_i$ is degree-preserving. 
We obtain \eqref{x_assump_scalar} by noting that 
\begin{equation}
    a^\dagger a = \sum_{m\geq 0} m\ketbra{m}{m}
\end{equation}
and therefore, on the degree-$k$ subspace $\mc{S}_k$, we have  
\begin{equation}
    \sum_i a_i^\dagger a_i \ket{\alpha_1\cdots \alpha_n} = \left(\sum_i \alpha_i\right) \ket{\alpha_1\cdots \alpha_n} = k \ket{\alpha_1\cdots \alpha_n}. 
\end{equation}
Thus we obtain \eqref{eqn:uds expectation} {for $\mc{L}_0$,} with $\gamma=\olg \deg\ket{\alpha}$. 

We further note that, since the dual-rail subspace is a subspace of $\mc{S}_k$ in the case $d=2$, the above also proves the UDS conditions are satisfied for that subspace. 


\subsubsection{Variant: unitary transformations}
We also consider the following more general variant. 
{
We express a general Fock basis vector in terms of creation operators, $\ket{\alpha_1 \cdots \alpha_n} = (a_1^\dagger)^{\alpha_1} \cdots (a_n^\dagger)^{\alpha_n}\ket{\vec{0}}$, 
where $\ket{\vec{0}} = \ket{0\cdots 0}$ is the vacuum state. 
For any $n\times n$ unitary matrix $U = (u_{ij})$, 
we consider a {degree or particle number preserving unitary transformation $\widetilde{U}$} with the property $\widetilde{U} a_j^\dagger \widetilde{U}^\dagger = \sum_{i} u_{ij} a_i^\dagger$. In photonics, these are precisely the linear optical unitaries, a proper subgroup of the group of all photon number-preserving unitaries. Additionally, notice that $\widetilde{U}\ket{\vec{0}} = \ket{\vec{0}}$ (since $\tilde U$ preserves particle number, it does not change the vacuum state). 
Explicitly, $\widetilde{U}$ satisfies}
\begin{align*}
    {\widetilde{U} \ket{\alpha_1 \cdots \alpha_n}} & {=\widetilde{U}(a_1^\dagger)^{\alpha_1} \cdots (a_n^\dagger)^{\alpha_n}\ket{\vec{0}}}
    \\&{= (\widetilde{U} a_1^\dagger \widetilde{U}^\dagger)^{\alpha_1}\cdots (\widetilde{U} a_n^\dagger \widetilde{U}^\dagger)^{\alpha_n}\widetilde{U}\ket{\vec{0}}}
    \\&{= (u_{11}a_1^\dagger + u_{12} a_2^\dagger + \dots + u_{1n} a_n^\dagger)^{\alpha_1}\cdots (u_{n1}a_1^\dagger + u_{n2} a_2^\dagger + \dots + u_{nn} a_n^\dagger)^{\alpha_n}\ket{\vec{0}},}
\end{align*}
{where $a_i^\dagger$ refers to the creation operator on the $i$th mode throughout.} 
Note that we have 
$
    {\widetilde{U}} a_j {\widetilde{U}}^\dagger = \sum_{i} u_{ij}^* {a_i}, 
$
a sum of annihilation operators on different qubits. 

We now consider the case of \eqref{x_assump_lindblad} with $\mc{B} = \{1, \dots, n\}$ and $A_i = {\widetilde{U}} a_i {\widetilde{U}}^\dagger$. 
We use the same subspaces $\mc{V}$, $\mc{S}_k$, etc., as in the previous example. 
Since $A_i$ is a sum of annihilation operators, it lowers the degree of Fock basis vectors by $1$, and \eqref{x_assump_scalar} follows as before. 
Note also that ${\widetilde{U}}$ preserves the degree and therefore leaves the computational subspace $\mc{S}_k$ of degree-$k$ vectors invariant. 
Then for \eqref{x_assump_scalar}, we have
\begin{equation}
    \sum_i A_i^\dagger A_i \ket{\alpha_1\cdots \alpha_n} = \sum_i {\widetilde{U}} a_i^\dagger {\widetilde{U}}^\dagger {\widetilde{U}} a_i {\widetilde{U}}^\dagger \ket{\alpha_1\cdots \alpha_n} = \sum_i {\widetilde{U}} a_i^\dagger a_i {\widetilde{U}}^\dagger \ket{\alpha_1\cdots \alpha_n} = \deg \alpha \ket{\alpha_1\cdots \alpha_n} = k\ket{\alpha_1\cdots \alpha_n}. 
\end{equation}
We note that this example behaves very similarly to the above, but now the operators $A_i$ are no longer required to correspond to single-qudit noise. 

\subsubsection{Correlated Decays}
Another potential example corresponds to correlated decays, as in the Lindblad operator 
\begin{equation*}
    \mc{L}\rho = \mc{H}\rho + \sum_{i\neq j} \gamma_{ij} {\mc{D}}[a_i a_j]\rho. 
\end{equation*}
Note that 
\begin{equation}
    \sum_{i\neq j}a_i^\dagger a_j^\dagger a_i a_j\ket{\alpha_1\cdots \alpha_n} = \sum_{i\neq j}\alpha_i \alpha_j \ket{\alpha_1\cdots \alpha_n}
    =\left(\left(\sum_i \alpha_i\right)^2 - \sum_i \alpha_i^2\right)\ket{\alpha_1\cdots \alpha_n}, 
\end{equation} 
so in order to satisfy \eqref{x_assump_scalar}, we must change our computational subspace to the span of Fock basis vectors with a fixed nonzero ``variance." 
(This distinction is irrelevant in the qubit case, since all $\alpha_i$ are $0$ or $1$.) 
This change gives \eqref{x_assump_scalar}, and \eqref{eq:invariant} is easily checked. 

Alternatively, one may consider 
\begin{equation*}
    \mc{L}\rho = \mc{H}\rho + \sum_{i,j} \gamma_{ij} {\mc{D}}[a_i a_j]\rho, 
\end{equation*}
where the restriction $i\neq j$ is removed. For $\ket{\alpha_1\cdots \alpha_n}$ in the degree-$k$ subspace, so that $\sum_i \alpha_i = k$, this satisfies 
\begin{equation}
    \sum_{i, j}a_i^\dagger a_j^\dagger a_i a_j\ket{\alpha_1\cdots \alpha_n} = \left(\left(\sum_i \alpha_i\right)^2 - \sum_i \alpha_i\right)\ket{\alpha_1\cdots \alpha_n} = (k^2 - k)\ket{\alpha_1\cdots \alpha_n}.
\end{equation} 
Then this setting may use the usual degree-$k$ computational subspace, and the UDS assumptions \eqref{eq:invariant}, \eqref{x_assump_scalar} easily follow. 
Since this example uses the same computational subspace as the standard amplitude damping case, one can actually show that they are compatible: with the Lindblad equation 
\begin{equation*}
    \mc{L}\rho = \mc{H}\rho + \sum_i \gamma_i {\mc{D}}[a_i]\rho + \sum_{i,j} \gamma_{ij} {\mc{D}}[a_i a_j]\rho,
\end{equation*}
one may define 
\begin{equation*}
    {\mc{D}_0'} = \left(\frac{1}{n}\sum_b \gamma_b\right)\sum_i {\mc{D}}[a_i]\rho + \left(\frac{1}{n^2}\sum_{b,d}\gamma_{b,d}\right)\sum_{i,j} {\mc{D}}[a_i a_j]\rho
\end{equation*}
and find that \eqref{eqn:uds expectation} holds for $\hmc{L}_0 = \hmc{H} + {\hmc{D}_0'}$. 

Similar considerations allow for transformations by a unitary $U$ as above and correlated decays between more than $2$ qudits. 
We will observe in Appendix~\ref{sec:damping examples cancellation}, however, that the examples based on correlated decays seem not to conveniently allow for cancellation of the first-order error. 

\subsection{Perturbative Cancellation}\label{sec:perturbative}

\subsubsection{Additional assumption and its consequences}

We now return to the notation and generality of Appendix~\ref{sec:explicit}, in particular \eqref{x_assump_lindblad}, \eqref{eq:invariant}, \eqref{x_assump_scalar}. 
In addition to these assumptions, we let $T$ be a unitary operator that commutes with $\mc{P}$, with corresponding superoperator $\mc{T}(\rho) = T\rho T^\dagger$. 
We assume there exists a permutation $\pi$ of the index set $\mc{B}$ such that
\begin{equation}\label{x_assump_t}
     TA_\beta  T^\dagger = A_{\pi(\beta)}\textnormal{ for all }\beta, 
\end{equation}
and that $\pi$ acts transitively on $\mc{B}$. 
As a consequence, note that $\hmc{T}$ acts on the operators ${\mc{D}}[A]$ similarly, with 
\begin{equation}\label{TDA}
    \hmc{T}{\hmc{D}}[A_\beta]\hmc{T}^{-1} = {\hmc{D}}[A_{\pi(\beta)}]. 
\end{equation}
For ${\hmc{D}_0}$ defined in \eqref{eq:def D}, this implies 
\begin{equation}\label{eq:TD}
    \hmc{T}{\hmc{D}_0}\hmc{T}^{-1} = {\hmc{D}_0}.
\end{equation}
For the amplitude damping case considered in Appendix~\ref{sec:main system}, 
we will take $ T$ to be the cyclic shift operator given by 
\begin{equation}
     T\ket{\alpha_1 \cdots \alpha_n} = \ket{\alpha_n \alpha_1 \alpha_2 \cdots \alpha_{n-1}}, 
\end{equation}
so that $\pi$ is the permutation $(1 \, 2 \,\cdots\, n)$ and we have $ T a_i  T^\dagger = a_{i+1}$ (indices taken modulo $n$). 

We now consider the perturbation by $\hmc{L}_1$ and show that the uniformly decaying subspace is stable to first order under varying decay rates. 
We note that the perturbation $\mc L_1$ induces a bias in the expectation value. 
We will account for this by averaging over multiple different experiments, shifting by a power of $ T$ each time. 
Let $j$ satisfy 
$0\leq j < m$, 
where $m=|\mc{B}|$ is the order of $ T$. 
We implement a $ T^j$-shifted version of the Hamiltonian, $\mc{T}^j (H) = T^j H T^{-j}$, which corresponds to replacing $\hmc{H}$ by $\hmc{T}^j \hmc{H} \hmc{T}^{-j}$ and $\hmc{L}_0$ by $\hmc{T}^j\hmc{H} \hmc{T}^{-j}+ {\hmc{D}_0}$. 
We similarly implement shifted versions of the state and observable, $ \mc{T}^j(\rho)$ and $\mc{T}^j(O)$ respectively. 
We observe that, by \eqref{eq:TD}, 
\begin{equation}
    \hmc{T}^{-j}(\hmc{T}^j\hmc{H} \hmc{T}^{-j}+ {\hmc{D}_0} + \hmc{L}_1)\hmc{T}^j = \hmc{H} + {\hmc{D}_0} + \hmc{T}^{-j}\hmc{L}_1\hmc{T}^{j} = \hmc{L}_0 + \hmc{T}^{-j}\hmc{L}_1\hmc{T}^{j}.
\end{equation}
The expected value of the $T^j$-shifted experiment is then  
\begin{align}
    \bbra{O} \hmc{T}^{-j}e^{t(\hmc{T}^j \hmc{H}\hmc{T}^{-j} + {\hmc{D}_0} + \hmc{L}_1)}\hmc{T}^j \kket{\rho(0)}
    &= \bbra{O} e^{t( \hmc{L}_0 + \hmc{T}^{-j}\hmc{L}_1\hmc{T}^{j})} \kket{\rho(0)},
\end{align}
the same as the unshifted case with $\hmc{L}_1$ replaced by $\hmc{T}^{-j}\hmc{L}_1\hmc{T}^{j}$. 
We note that by \eqref{TDA} and the definition of $\hmc{L}_1$, 
\begin{align}\label{eq:TLT}
    \hmc{T}^{-j}\hmc{L}_1\hmc{T}^{j} = \sum_\beta \Delta\gamma_\beta \hmc{T}^{-j}{\hmc{D}}[A_\beta]\hmc{T}^{j} = \sum_\beta \Delta\gamma_\beta {\hmc{D}}[A_{\pi^{-j}(\beta)}] = \sum_\beta \Delta\gamma_{\pi^j(\beta)} {\hmc{D}}[A_{\beta}].
\end{align}
In particular, recalling that 
$\pi$ transitively permutes the indices $\beta\in\mc{B}$ and that 
\begin{equation*}
    \sum_\beta \Delta\gamma_\beta = \sum_\beta\left( \gamma_\beta - \olg \right) = 
    \sum_\beta\gamma_\beta - m\olg = 0,
\end{equation*}
we observe that
\begin{equation}\label{L avg}
    \sum_{j=0}^{m-1} \hmc{T}^{-j}\hmc{L}_1\hmc{T}^{j} = \sum_\beta \left(\sum_{j=0}^{m-1}\Delta\gamma_{\pi^j(\beta)}\right) {\hmc{D}}[A_{\beta}] = \sum_\beta \left(\sum_{\beta'\in\mc{B}}\Delta\gamma_{\beta'}\right) {\hmc{D}}[A_{\beta}]
    =0.
\end{equation}

\subsubsection{First order error cancellation}

Let us briefly return to the unshifted case. We work in the Liouville space 
\begin{align}
    \kket{\dot \rho(t)} = (\hat{\mc L}_0 + \hat{\mc L}_1) \kket{\rho(t)}.
\end{align}
We consider the interaction picture where we define $\kket{\rho_I(t)} \equiv e^{-\hat{\mc L}_0 t} \kket{\rho(t)}$ and $\hat{\mc L}_{1}^I(t) \equiv e^{-\hat{\mc L}_0 t} \hat{\mc L}_1 e^{\hat{\mc L}_0 t}$ and subsequently obtain the equation of motion 
\begin{align}
     \kket{\dot \rho_I(t)} = \hat{\mc L}_{1}^I (t) \kket{ \rho_I(t)} \;.
\end{align}
We can now perturbatively solve the above equation as 
\begin{align}
\kket{\rho_I(t)} = \kket{\rho(0)} + \int_{0}^t dt^\prime \hat{\mc L}_1^I(t^\prime) \kket{\rho(0)} +  
\int_{0}^t dt^\prime \int_{0}^{t^{\prime}} dt^{\prime\prime} \hat{\mc L}_1^I(t^\prime)  \hat{\mc L}_1^I(t^{\prime\prime}) \kket{\rho_I(t'')}\;.
\end{align}
where we used $\kket{\rho_I(0)} = \kket{\rho(0)}$. 
Applying $e^{\hat{\mc L}_0 t}$, we obtain
\begin{align}\label{twice iterated}
\kket{\rho(t)} = e^{\hat{\mc L}_0 t}\kket{\rho(0)} + \int_{0}^t dt^\prime e^{\hat{\mc L}_0 t} \hat{\mc L}_1^I(t^\prime) \kket{\rho(0)} +  
\int_{0}^t dt^\prime \int_{0}^{t^{\prime}} dt^{\prime\prime} e^{\hat{\mc L}_0 t}\hat{\mc L}_1^I(t^\prime)  \hat{\mc L}_1^I(t^{\prime\prime}) \kket{\rho_I(t'')}\;.
\end{align}
Therefore the expectation value of the observable $O$ now is 
\begin{align}
    \langle O \rangle_{\text{UDS}} = \bbra{O}{\rho(t)}\rrangle &= \llangle O| e^{\hat{\mc L}_0 t} \bigg( \kket{\rho(0)} + \int_{0}^t dt^\prime \hat{\mc L}_1^I(t^\prime) \kket{\rho(0)} +  \int_{0}^t dt^\prime \int_{0}^{t^{\prime}} dt^{\prime\prime} e^{\hat{\mc L}_0 t}\hat{\mc L}_1^I(t^\prime)  \hat{\mc L}_1^I(t^{\prime\prime}) \kket{\rho_I(t'')} \bigg)\;
    \\&= e^{-c\bar\gamma t} \langle O \rangle_C + \llangle O| e^{\hat{\mc L}_0 t} \bigg( \int_{0}^t dt^\prime \hat{\mc L}_1^I(t^\prime) \kket{\rho(0)} +  \int_{0}^t dt^\prime \int_{0}^{t^{\prime}} dt^{\prime\prime} e^{\hat{\mc L}_0 t}\hat{\mc L}_1^I(t^\prime)  \hat{\mc L}_1^I(t^{\prime\prime}) \kket{\rho_I(t'')} \bigg)\;,\label{2nd order expansion}
\end{align}
where we apply \eqref{eqn:uds expectation} in the last line. 
We will consider the $0$th and $1$st order terms, observing that averaging over all $\hmc{T}$-shifts leaves the $0$th order invariant while cancelling the $1$st order terms. 
We will consider the problem of bounding the $2$nd order terms in Appendix~\ref{sec:second order}. 

Since $\hmc{L}_0$ is unaffected by $\hmc{T}$-shifting (up to the change in encoding of the Hamiltonian), the constant term $e^{-c\bar\gamma t} \langle O \rangle_C$ is unaffected. 
Also recall $\hat{\mc L}_{1}^I(t) \equiv e^{-\hat{\mc L}_0 t} \hat{\mc L}_1 e^{\hat{\mc L}_0 t}$, so the $\hmc{T}^j$-shifted version is 
$e^{-\hat{\mc L}_0 t} \hmc{T}^{-j}\hat{\mc L}_1 \hmc{T}^j e^{\hat{\mc L}_0 t}$. 
Then the expected value averaged over all $\hmc{T}$-shifts, denoted by $\llangle O \rrangle_{\text{UDS}}$, satisfies 
\begin{align}
    \llangle O \rrangle_{\text{UDS}} - e^{-c\bar\gamma t} \langle O \rangle_C 
    &=  \frac{1}{m}\sum_{j=0}^{m-1} \llangle O| e^{\hat{\mc L}_0 t} \int_{0}^t dt^\prime e^{-\hat{\mc L}_0 t} \hmc{T}^{-j}\hat{\mc L}_1 \hmc{T}^j e^{\hat{\mc L}_0 t} \kket{\rho(0)} + \textnormal{2nd order}\;
    \\&= \frac{1}{m} \llangle O| e^{\hat{\mc L}_0 t} \int_{0}^t dt^\prime e^{-\hat{\mc L}_0 t} \left(\sum_{j=0}^{m-1}\hmc{T}^{-j}\hat{\mc L}_1 \hmc{T}^j\right) e^{\hat{\mc L}_0 t} \kket{\rho(0)} + \textnormal{2nd order}\;
    \\&= 0 + \textnormal{2nd order},
\end{align}
where we apply \eqref{L avg} in the last line. 
In particular, we were able to negate the effect of first order deviations in decay rates even without the knowledge of perturbations. Therefore, we see that uniformly decaying subspaces can be made stable up to first order.

\subsubsection{Amplitude damping channel examples}\label{sec:damping examples cancellation}
We now return to the examples related to the amplitude damping channel, discussed in Appendix~\ref{sec:main system}. 
As stated above, in the standard case $A_i = a_i$ we take $ T$ to be the cyclic shift operator given by 
\begin{equation}
     T\ket{\alpha_1 \cdots \alpha_n} = \ket{\alpha_n \alpha_1 \alpha_2 \cdots \alpha_{n-1}}, 
\end{equation}
so that $ T a_i  T^\dagger = a_{i+1}$ (indices taken modulo $n$). 

More generally, we recall the variant with $A_i = {\widetilde{U}} a_i {\widetilde{U}}^\dagger$ where $U$ is an $n\times n$ unitary matrix. 
We note that 
\begin{equation*}
    ({\widetilde{U}} T{\widetilde{U}}^\dagger) A_i ({\widetilde{U}} T {\widetilde{U}}^\dagger)^\dagger = {\widetilde{U}}  T a_i  T^\dagger {\widetilde{U}}^\dagger = {\widetilde{U}} a_{i+1} {\widetilde{U}}^\dagger
    =A_{i+1},
\end{equation*}
where we continue to take the indices of the $A$ and $a$ operators modulo $n$. 
Then in the ${\widetilde{U}}$-transformed setting, we must replace the translation $ T$ with  ${\widetilde{U}} T{\widetilde{U}}^\dagger$. 

For the case of correlated decays, $A_{ij} = a_i a_j$, there is in general no operator $ T$ satisfying the desired properties. 
In principle, one can simply replace the role of $1,  T,  T^2, \dots$ with the entire group of permutations of $n$ qudits, $S_n$, 
and similar calculations to the above will work. 
But of course this would require one to perform $n!$ different experiments, leading to infeasible scaling. 

\subsubsection{Second order error bound}\label{sec:second order}



We now consider the last term in \eqref{2nd order expansion}. 
We have shown that under the assumptions of the previous section, the middle (first order) term can be cancelled. 
Thus if one wants to estimate $\langle O \rangle_C\approx e^{c\bar\gamma t}\llangle O \rrangle_{\text{UDS}}$, 
the error is given by the last term of \eqref{2nd order expansion}. 
To obtain this bound, we follow the techniques of the supplemental material of \cite{temme2017error}. 

Since $\bbra{O} = \bbra{O}\hat{\mc{P}}$, $\hat{\mc{P}} \hmc{L}_i = \hat{\mc{P}} \hmc{L}_i\hat{\mc{P}}$, and $\hmc{P}\hat{\mc L}_0\hmc{P} = \hmc{P}\hmc{H}\hmc{P} - c\olg \hmc{P}$, we have
\begin{align}
    \bbra{O}  e^{\hat{\mc L}_0 t}\hat{\mc L}_1^I(t^\prime)  \hat{\mc L}_1^I(t^{\prime\prime}) \kket{\rho_I(t'')}
    &= \bbra{O}  \hmc{P}e^{\hat{\mc L}_0 t}\hat{\mc L}_1^I(t^\prime)  \hat{\mc L}_1^I(t^{\prime\prime}) \kket{\rho_I(t'')}
    \\ &= \bbra{O}  e^{\hmc{P}\hat{\mc L}_0\hmc{P} t}\hmc{P}\hat{\mc L}_1^I(t^\prime)\hmc{P}  \hat{\mc L}_1^I(t^{\prime\prime})\hmc{P} \kket{\rho_I(t'')}
    \\&= e^{-c\olg t}\bbra{O}  e^{\hmc{P}\hmc{H}\hmc{P} t}\hmc{P}\hat{\mc L}_1^I(t^\prime)\hmc{P}  \hat{\mc L}_1^I(t^{\prime\prime})\hmc{P} \kket{\rho_I(t'')}.
\end{align}
We now apply the mean value theorem: there exist $\zeta', \zeta^{\prime\prime}\in [0,t]$ such that 
\begin{equation}
    \int_{0}^t dt^\prime \int_{0}^{t^{\prime}} dt^{\prime\prime} e^{-c\olg t}\bbra{O}  e^{\hmc{P}\hmc{H}\hmc{P} t}\hmc{P}\hat{\mc L}_1^I(t^\prime)\hmc{P}  \hat{\mc L}_1^I(t^{\prime\prime})\hmc{P} \kket{\rho_I(t'')} 
    = \dfrac{t^2}{2}e^{-c\olg t}\bbra{O}  e^{\hmc{P}\hmc{H}\hmc{P} t}\hmc{P}\hat{\mc L}_1^I(\zeta^\prime)\hmc{P}  \hat{\mc L}_1^I(\zeta^{\prime\prime})\hmc{P} \kket{\rho_I(\zeta'')}.
\end{equation}
We briefly recall the Schatten $p$-norms, for $1\leq p\leq \infty$, $||T||_p = \left(\Tr( \sqrt{T^\dagger T}^p ) \right)^{1/p}$. 
For $p=\infty$, this is the standard operator norm. Rewriting the above as a $1$-norm and applying H\"{o}lder's inequality for Schatten $p$-norms, we have 
\begin{align}
    \dfrac{t^2}{2}e^{-c\olg t}\left|\bbra{O}  e^{\hmc{P}\hmc{H}\hmc{P} t}\hmc{P}\hat{\mc L}_1^I(\zeta^\prime)\hmc{P}  \hat{\mc L}_1^I(\zeta^{\prime\prime})\hmc{P} \kket{\rho_I(\zeta'')}\right|
    &= \dfrac{t^2}{2}e^{-c\olg t}||\bbra{O}  e^{\hmc{P}\hmc{H}\hmc{P} t}\hmc{P}\hat{\mc L}_1^I(\zeta^\prime)\hmc{P}  \hat{\mc L}_1^I(\zeta^{\prime\prime})\hmc{P} \kket{\rho_I(\zeta'')}||_1
    \\&\leq \dfrac{t^2}{2} e^{-c\olg t}||\bbra{O}  e^{\hmc{P}\hmc{H}\hmc{P} t}||_\infty \cdot || \hmc{P}\hat{\mc L}_1^I(\zeta^\prime)\hmc{P}  \hat{\mc L}_1^I(\zeta^{\prime\prime})\hmc{P} \kket{\rho_I(\zeta'')}||_1.
\end{align}
Since $e^{\hmc{P}\hmc{H}\hmc{P}t}$ is unitary, we have $||\bbra{O}  e^{\hmc{P}\hmc{H}\hmc{P} t}||_\infty = ||O||_\infty = ||O||_{op}$, the standard operator norm of $O$. 
Then we must only bound 
\begin{align}
    || \hmc{P}\hat{\mc L}_1^I(\zeta^\prime)\hmc{P}  \hat{\mc L}_1^I(\zeta^{\prime\prime})\hmc{P} \kket{\rho_I(\zeta'')}||_1 &= || \hmc{P}\hat{\mc L}_1^I(\zeta^\prime) \hmc{P}  \hat{\mc L}_1^I(\zeta^{\prime\prime}) \hmc{P}\kket{\rho_I(\zeta^{\prime\prime})}  ||_1
    \\&= ||e^{-\hmc{P}\hat{\mc L}_0\hmc{P} \zeta^\prime} \hmc{P}\hat{\mc L}_1\hmc{P} e^{\hmc{P}\hat{\mc L}_0\hmc{P}(\zeta^\prime - \zeta^{\prime\prime})} \hmc{P}\hat{\mc L}_1\hmc{P}\kket{\rho(t'')}  ||_1
    \\&= e^{-c\olg (-\zeta^\prime + (\zeta^\prime - \zeta^{\prime\prime})} 
    || e^{-\hmc{P}\hmc{H}\hmc{P} \zeta^\prime} \hmc{P}\hat{\mc L}_1\hmc{P} e^{\hmc{P}\hmc{H}\hmc{P}(\zeta^\prime - \zeta^{\prime\prime})} \hmc{P}\hat{\mc L}_1\hmc{P}\kket{\rho(\zeta^{\prime\prime})} ||_1
    \\&\leq e^{c\olg \zeta^{\prime\prime}} 
    || e^{-\hmc{P}\hmc{H}\hmc{P} \zeta^\prime} \hmc{P}\hat{\mc L}_1\hmc{P} e^{\hmc{P}\hmc{H}\hmc{P}(\zeta^\prime - \zeta^{\prime\prime})} \hmc{P}\hat{\mc L}_1\hmc{P}||_{op,1}
    \\&= e^{c\olg \zeta^{\prime\prime}} 
    || \hmc{P}\hat{\mc L}_1\hmc{P}||_{op,1}^2,
\end{align}
where $||\cdot||_{op,1}$ is the operator norm with respect to the Schatten $1$-norm, and we use the fact that operator norms are unitarily invariant. 
In total, then, the error term of interest satisfies
\begin{align}
    \left|\int_{0}^t dt^\prime \int_{0}^{t^{\prime}} dt^{\prime\prime} e^{-c\olg t}\bbra{O}  e^{\hmc{P}\hmc{H}\hmc{P} t}\hmc{P}\hat{\mc L}_1^I(t^\prime)\hmc{P}  \hat{\mc L}_1^I(t^{\prime\prime})\hmc{P} \kket{\rho_I(t'')} \right| 
    &\leq \dfrac{t^2}{2}e^{-c\olg(t-\zeta^{\prime\prime})} ||O||_{op} \cdot || \hmc{P}\hat{\mc L}_1\hmc{P}||_{op,1}^2
    \\&\leq \dfrac{t^2}{2}||O||_{op} \cdot || \hmc{P}\hat{\mc L}_1\hmc{P}||_{op,1}^2. 
\end{align}
The analogous result holds in the $ T$-shifted case. Combining this with the results of Appendix~\ref{sec:perturbative}, we obtain
\begin{equation}
    \Big|\llangle O \rrangle_{\text{UDS}} - e^{-c\bar\gamma t} \langle O \rangle_C\Big| \leq \dfrac{t^2}{2}||O||_{op} \cdot || \hmc{P}\hat{\mc L}_1\hmc{P}||_{op,1}^2. 
\end{equation}

\subsubsection{Return to amplitude damping example}

We return to the example of Appendix~\ref{sec:main system}, with qudits of dimension $d$, $A_i = a_i$ the annihilation operator on the $i$th qudit, and computational subspace the degree-$k$ subspace $\mc{S}_k$. 
The bound above is determined by the operator norm of $\hmc{P}\hat{\mc L}_1\hmc{P}$. 
We note
\begin{equation*}
    \mc{P}{\mc L}_1\mc{P}(\rho) =  P\mc{L}_1( P\rho P) P = -\frac{1}{2}\sum_i \Delta\gamma_i \{ a_i^\dagger a_i, \rho\} = -\frac{1}{2}\sum_i \Delta\gamma_i \left\{ \sum_s s\ketbra{s}, \rho\right\}.
\end{equation*}
In particular, $\hmc{P}\hat{\mc L}_1\hmc{P}$ is diagonalizable in the Fock basis $\ketbra{\alpha_1 \cdots \alpha_n}{\alpha'_1\cdots \alpha'_n}$, 
so its operator norm is the modulus of the largest eigenvalue. 
One can easily verify that this largest eigenvalue is bounded above by $k\lambda$, where $\lambda= \max_i |\Delta\gamma_i|$. 
This gives the bound
\begin{equation}
    \Big|\llangle O \rrangle_{\text{UDS}} - e^{-k\bar\gamma t} \langle O \rangle_C\Big| \leq \dfrac{(tk\lambda)^2}{2}||O||_{op}.
\end{equation}
The same result holds in the case $A_i = {\widetilde{U}} a_i {\widetilde{U}}^\dagger$. 

\subsection{Circuit case}\label{sec:circuit}

We now discuss why one can apply the techniques discussed above, specifically the cancellation of errors up to first order, to the discretized example at the end of the main text.  

We let $\mc{E} = \mc{E}(\eta)$ be a single-qudit discretized amplitude damping channel, given by the following Kraus operators $K_s(\eta)$ for $s\geq 0$ and a parameter $\eta$ \cite{chuang1997bosonic} (where $\eta = 1-\gamma$ in ibid.):
\begin{equation}
    K_s(\eta) = \sum_r \sqrt{\binom{r}{s} \eta^{r-s}(1-\eta)^s}\ketbra{r-s}{r}.
\end{equation}
We write $\mc{E}_i(\eta_i)$ for $\mc{E}$ acting on the $i$th qudit with parameter $\eta = \eta_i$. 

Now suppose we prepare a state $\rho\in\omc{V}$, apply unitaries $U_1, \dots, U_{N_T}$ on $\mc{V}$ that commute with $P$ (with corresponding operators $\hmc{U}_1, \dots, \hmc{U}_{N_T}$ on the Liouville space), and measure an observable $O$ with $ PO P = O$. 
We model dissipation in a discretized manner by replacing each $\hmc{U}_j$ by $(\bigotimes_i \hmc{E}_i(\eta_i))\hmc{U}_j$. 
The expectation value is the following: 
\begin{equation}
    \langle O \rangle = \bbra{O}\hmc{P}(\bigotimes_i \hmc{E}_i(\eta_i))\hmc{U}_1 (\bigotimes_i \hmc{E}_i(\eta_i))\cdots (\bigotimes_i \hmc{E}_i(\eta_i)) \hmc{U}_{N_T} \kket{\rho}.
\end{equation}
Since all $K_s$ preserve or lower the degree, keeping $\omc{S}^\perp$ and $\omc{V}$ invariant, we have $\hmc{P} \hmc{E}_i = \hmc{P} \hmc{E}_i\hmc{P}$. Further, the $\hmc{U}_i$ commute with $\hmc{P}$ by assumption. Therefore, 
\begin{equation}
    \langle O \rangle = \bbra{O}(\bigotimes_i \hmc{P}\hmc{E}_i(\eta_i)\hmc{P})\hmc{U}_1 (\bigotimes_i \hmc{P}\hmc{E}_i(\eta_i)\hmc{P})\cdots (\bigotimes_i \hmc{P}\hmc{E}_i(\eta_i)\hmc{P}) \hmc{U}_{N_T} \kket{\rho}.
\end{equation}
Since $ POP = O$ (meaning that we post-select for measurements in the degree-$k$ subspace), and $\kket{\rho}$ involves only terms of degree $k$ or lower, all terms involving the degree-raising operators $K_s$ for $s>0$ must vanish. Then only the $K_0$ term of each $\hmc{P}\hmc{E}_i\hmc{P}$ contributes to the expectation. 
Letting $\hat{K}_0^{(i)}(\eta)$ be the corresponding operator on $\omc{V}$, we note that we may replace $\bigotimes_i \hmc{P}\hmc{E}_i(\eta_i)\hmc{P}$ with 
\begin{equation}
    \hms{E}(\eta_1, \dots, \eta_n)\equiv \bigotimes_i \hmc{P}\hat{K}_0^{(i)}(\eta_i)\hmc{P}, 
\end{equation}
and write
\begin{equation}
    \langle O \rangle = \bbra{O}\hms{E}(\eta_1, \dots, \eta_n)\hmc{U}_1 \hms{E}(\eta_1, \dots, \eta_n)\cdots \hms{E}(\eta_1, \dots, \eta_n) \hmc{U}_{N_T} \kket{\rho}.
\end{equation}

We will show that the first-order cancellation observed above extends to this discretized setting. 
Toward that end, for comparison with the continuous case, 
we take $\eta_i = \exp(-\Delta t \gamma_i)$, where $\gamma_i$ is the parameter in the Lindblad master equation \eqref{lindblad damping} and $\Delta t = t/{N_T}$ is the time elapsed per unitary (here $t$ is the total time elapsed, as above). 
We define 
\begin{equation}
    \overline{\eta} = (\eta_1\cdots \eta_n)^{1/n} = \exp(-\frac{1}{n}\Delta t \sum_i \gamma_i) = \exp(-\Delta t \olg)\textnormal{ and }\Delta\eta_i = \eta_i/\overline{\eta} = \exp(-\Delta t \Delta\gamma_i),
\end{equation}
and note that $\hat{K}_0^{(i)}(\eta_i) = \hat{K}_0^{(i)}(\overline{\eta})\hat{K}_0^{(i)}(\Delta\eta_i)$, so that
\begin{equation}
    \hms{E}(\eta_1, \dots, \eta_n) = \hms{E}(\overline{\eta}, \dots, \overline{\eta})\hms{E}(\Delta\eta_1, \dots, \Delta\eta_n).
\end{equation}
For $\kket{\rho'}$ in the degree-$k$ subspace, we note that
\begin{equation}
    \hms{E}(\overline{\eta}, \dots, \overline{\eta})\kket{\rho'} = \overline{\eta}^k = \exp(-k\Delta t\olg),
\end{equation}
so we obtain
\begin{equation}
    \langle O \rangle = e^{-k{N_T}\Delta t\olg}\bbra{O}\hms{E}(\Delta\eta_1, \dots, \Delta\eta_n)\hmc{U}_1 \hms{E}(\Delta\eta_1, \dots, \Delta\eta_n)\cdots \hms{E}(\Delta\eta_1, \dots, \Delta\eta_n) \hmc{U}_{N_T} \kket{\rho},
\end{equation}
where we note that ${N_T}\Delta t = t$ is the total time. 

Now, we consider a single expression 
$\hms{E}(\Delta\eta_1, \dots, \Delta\eta_n)\kket{\rho'}$ for $\rho'\in\omc{V}$. 
As above, it suffices to consider $\rho' \in\omc{S}$. 
For $\rho' = \ketbra{\alpha_1\cdots \alpha_n}{\alpha'_1\cdots \alpha'_n}\in\omc{S}$, $\hmc{P}\hmc{E}_i(\Delta\eta_i)\hmc{P}$ acts by the scalar $(\Delta\eta_i)^{(\alpha_i + \alpha'_i)/2}.$ 
In particular, we see that $\hms{E}(\Delta\eta_1, \dots, \Delta\eta_n)$ acts by the scalar
\begin{equation}
    \prod_i (\Delta\eta_i)^{(\alpha_i + \alpha'_i)/2} = \exp(-\frac{1}{2}\Delta t \sum_i (\alpha_i+\alpha'_i)\Delta\gamma_i) = 1 -\frac{1}{2}\Delta t \sum_i (\alpha_i+\alpha'_i)\Delta\gamma_i + O((\Delta t \Delta\gamma_{\textnormal{max}})^2).
\end{equation}
Averaging over all cyclic permutations of the qudits as in Appendix~\ref{sec:perturbative}, we see that the averaged channel acts on $\rho' = \ketbra{\alpha_1\cdots \alpha_n}{\alpha'_1\cdots \alpha'_n}\in\omc{S}$ by the scalar 
\begin{equation}
    1 -\frac{1}{2n}\Delta t \sum_i (\alpha_i+\alpha'_i)\sum_j\Delta\gamma_j + O((\Delta t \Delta\gamma_{\textnormal{max}})^2) = 1 + O((\Delta t \Delta\gamma_{\textnormal{max}})^2).
\end{equation}
Then up to first order in $\Delta t$ and $\Delta\gamma_{\textnormal{max}}$, the superoperator $\hms{E}(\Delta\eta_1, \dots, \Delta\eta_n)$ acts trivially. 
This suffices to show that 
\begin{equation}
    \llangle O \rrangle_{\text{UDS}} = e^{-kt\olg}\left(\bbra{O}\hmc{U}_1 \cdots \hmc{U}_{N_T} \kket{\rho} + O((\Delta t \Delta\gamma_{\textnormal{max}})^2) \right)
\end{equation}
and
\begin{equation}
    \llangle O \rrangle_{\text{UDS}} - e^{-c\bar\gamma t} \langle O \rangle_C = O((\Delta t \Delta\gamma_{\textnormal{max}})^2)
\end{equation}
as desired.

\end{document}